\documentclass[aps,prc,twocolumn,showpacs,floatfix,tightenlines,superscriptaddress]{revtex4-1}

\usepackage{amsmath,amssymb,graphicx,bm}

\graphicspath{{figures/}}

\newcommand{\beqn}{\begin{equation}}
\newcommand{\eeqn}{\end{equation}}
\newcommand{\bea}{\begin{eqnarray}}
\newcommand{\eea}{\end{eqnarray}}

\newcommand{\abinit}{{\it ab initio }}

\newcommand{\mev}{\, \text{MeV}}

\newcommand{\nmax}{\ensuremath{N_{\rm max}}}
\newcommand{\hw}{\ensuremath{\hbar\Omega}}

\newcommand{\NNLO}{N$^2$LO}
\newcommand{\NNNLO}{N$^3$LO}

\begin{document}


\title{$^{12}$C properties with evolved chiral three-nucleon interactions}

\author{P.\ Maris}
\email{pmaris@iastate.edu}
\affiliation{Department of Physics and Astronomy, Iowa State University, 
Ames, IA\ 50011, USA}

\author{J.P.\ Vary}
\email{jvary@iastate.edu}
\affiliation{Department of Physics and Astronomy, Iowa State University, 
Ames, IA\ 50011, USA}

\author{A.\ Calci}
\email{angelo.calci@physik.tu-darmstadt.de}
\affiliation{Institut f\"ur Kernphysik, Technische Universit\"at Darmstadt, 64289 Darmstadt, Germany}

\author{J.\ Langhammer}
\email{joachim.langhammer@physik.tu-darmstadt.de}
\affiliation{Institut f\"ur Kernphysik, Technische Universit\"at Darmstadt, 64289 Darmstadt, Germany}

\author{S.\ Binder}
\email{sven.binder@physik.tu-darmstadt.de}
\affiliation{Institut f\"ur Kernphysik, Technische Universit\"at Darmstadt, 64289 Darmstadt, Germany}

\author{R.\ Roth}
\email{robert.roth@physik.tu-darmstadt.de}
\affiliation{Institut f\"ur Kernphysik, Technische Universit\"at Darmstadt, 64289 Darmstadt, Germany}


\begin{abstract}
We investigate selected static and transition properties of $^{12}$C
using \abinit No-Core Shell Model (NCSM) methods with chiral two- and
three-nucleon interactions.  We adopt the Similarity Renormalization
Group (SRG) to assist convergence including up to three-nucleon (3N)
contributions.  We examine the dependences of the $^{12}$C observables
on the SRG evolution scale and on the model-space parameters.  We
obtain nearly converged low-lying excitation spectra.  We compare
results of the full NCSM with the Importance Truncated NCSM in large
model spaces for benchmarking purposes. We highlight the effects of
the chiral 3N interaction on several spectroscopic observables. The
agreement of some observables with experiment is improved
significantly by the inclusion of 3N interactions, e.g., the B(M1)
from the first $J^{\pi}T=1^+1$ state to the ground state. However, in
some cases the agreement deteriorates, e.g., for the excitation energy
of the first $1^+0$ state, leaving room for improved next-generation
chiral Hamiltonians.
\end{abstract}

\date{\today}


\maketitle

\section{Introduction
\label{sec:introduction}}

No-Core Configuration Interaction methods have advanced rapidly in
recent years to make it feasible to accurately solve fundamental
problems in nuclear structure and reaction physics (e.g., see
Refs.~\cite{Navratil:2000ww,Navratil:2000gs,Barrett:2013nh,Maris:2008ax,Navratil:2009ut,Maris:2009bx,Roth:2009cw,Roth:2011ar,Roth:2011vt,Maris:2011as,Navratil:2011zs}).
At the same time, significant theoretical advances regarding the
underlying Hamiltonians, constructed within chiral effective field
theory (EFT), provide a foundation for nuclear many-body calculations
rooted in QCD \cite{EpHa09,MaEn11}.  In order to improve the
convergence behavior of the many-body calculations we employ a
consistent unitary transformation of the chiral Hamiltonians. Here we
use the Similarity Renormalization Group
(SRG)~\cite{Glazek:1993rc,Wegner:1994,Bogner:2007rx,Bogner:2009bt,Furnstahl:2012fn}
approach that provides a straightforward and flexible framework for
consistently evolving (softening) the Hamiltonian and other operators,
including three-nucleon
interactions~\cite{Jurgenson:2009qs,Roth:2011ar,Jurgenson:2013yya,RoCa13}.

The goal of this paper is twofold. First, we aim to provide results
for $^{12}$C spectra and other observables using realistic chiral
nucleon-nucleon (NN) plus three-nucleon (3N) interactions with
uncertainty estimates where feasible.  Second, we provide benchmark
comparisons between the full No-Core Shell Model
(NCSM)~\cite{Navratil:2000ww,Navratil:2000gs,Barrett:2013nh} and the
Importance Truncated No-Core Shell Model
(IT-NCSM)~\cite{Roth:2009cw,Roth:2011ar,Roth:2011vt,RoCa13}.

Previous investigations of $^{12}$C with chiral NN+3N interactions,
softened with the SRG approach, have mainly focused on the
ground-state energy~\cite{Roth:2011ar, Jurgenson:2013yya} and its
convergence properties. One of the directions in which the present
work extends these earlier efforts is by investigating a wider set of
observables including selected electromagnetic transitions and the
lowest-lying negative parity states. Our initial results for the
$2^+$ and $4^+$ rotational excited states were presented in
Ref. \cite{Maris:2012ccp}.

We limit our investigations to a single form of the chiral NN+3N
interaction. We use the chiral NN interaction at \NNNLO\ with 500\,MeV/c
cutoff from Ref.~\cite{Entem:2003ft} together with the 3N potential at
\NNLO~\cite{Epelbaum:2002vt} in the local form of
Ref.~\cite{Navratil:2007zn} with 500\,MeV/c cutoff and low-energy
constants determined entirely in the three-nucleon sector
\cite{GaQu09}. This is also the Hamiltonian used in
Refs.~\cite{Jurgenson:2009qs,Jurgenson:2010wy,Roth:2011ar,Jurgenson:2013yya,RoCa13}.
We evolve this Hamiltonian using the free-space SRG to
three representative flow parameters or momentum scales to examine the
scale-dependence of our results. As in the earlier applications, we
retain the induced many-body interaction through the
three-nucleon level and neglect induced four- and multi-nucleon
interactions.

In Section~\ref{sec:back}, we briefly review the formalism and
summarize related results from previous work.  The results for 
selected $^{12}$C observables are presented in
Section~\ref{sec:results}.  Section~\ref{sec:comparisons} presents
benchmarks of the IT-NCSM and NCSM. 
Finally, Section~\ref{sec:conclusions} summarizes
our conclusions and provides perspectives on future efforts.

\section{Theoretical Background
\label{sec:back}}

\subsection{NCSM and IT-NCSM}

We employ two related \abinit methods to solve for the properties of
$^{12}$C.  In the first approach, the NCSM, we follow
Refs.~\cite{Navratil:2000ww,Navratil:2000gs,Barrett:2013nh} where, for
a chosen NN and 3N interaction (either without or with SRG evolution)
we diagonalize the resulting many-body Hamiltonian in a sequence of
truncated harmonic-oscillator (HO) basis spaces. The basis spaces are
characterized by two parameters: $N_{\max}$ specifies the maximum
number of total HO quanta beyond the HO Slater determinant with all
nucleons occupying their lowest-allowed orbitals and $\hbar\Omega$
specifies the HO energy. The goal is to achieve convergence as
indicated by independence of these two basis parameters, either
directly or by extrapolation~\cite{Maris:2008ax}.

In the second approach, the IT-NCSM, we follow
Refs.~\cite{Roth:2009cw,Roth:2011ar,Roth:2011vt,RoCa13} where
subspaces of the $N_{\max}$-truncated spaces are dynamically selected
according to a measure derived from perturbation theory.  The IT-NCSM
uses this derived importance measure $\kappa_\nu$ for the individual
many-body basis states and retains only states with $|\kappa_\nu|$
above a threshold $\kappa_{\min}$ in the model space.  Through a
variation of this threshold and an {\it a posteriori} extrapolation
$\kappa_{\min}\rightarrow 0$ the contribution of discarded states is
recovered. We use the sequential update scheme discussed in
Refs.~\cite{Roth:2009cw,RoCa13}, which connects to the full NCSM model
space and, thus, to the exact NCSM results in the limit of vanishing
threshold. In the following we report threshold-extrapolated results
of the IT-NCSM at each $N_{\max}$ including an estimate for the
extrapolation uncertainties. In addition, we compare the IT-NCSM
results with the NCSM results in model spaces where we valuate
results from both approaches.

\subsection{Chiral NN+3N Interactions}

Chiral EFT has developed into a standard approach for the construction
of NN and 3N interactions with low-energy constants (LECs) fitted to
NN and 3N data. As mentioned above, we adopt the chiral EFT potential
at \NNNLO\ with 500\,MeV/c cutoff from Ref.~\cite{Entem:2003ft} together
with an 3N potential at \NNLO~\cite{Epelbaum:2002vt} in the local form
of Ref.~\cite{Navratil:2007zn} as this Hamiltonian was adopted for a
range of \abinit calculations of light and medium-mass nuclei and,
in particular, was used in previous works for $^{12}$C. For the LECs
introduced by the 3N interaction at \NNLO, we adopt the values fitted
to the $A=3$ binding energies and tritium
half-life~\cite{Gazit:2008ma}. That is, we adopt $c_D=-0.2$ and
$c_E=-0.205$ for a cutoff of 500\,MeV/c.

The first paper to report results for $^{12}$C with chiral NN+3N
interactions (with a different choice for $c_D$ and $c_E$) is
Ref.~\cite{Navratil:2007we}. That work employed the NCSM with the
Okubo-Lee-Suzuki (OLS) transformation
method~\cite{Okubo:1954zz,Suzuki:1980yp} to improve convergence and
presented natural parity results up through $N_{\max}=6$ basis
spaces. We considerably extend this span of basis spaces with the
present work and include the lowest unnatural parity states. Moreover,
we use the SRG evolution to soften the interaction instead of the OLS
transformation.
The SRG-evolved chiral NN+3N Hamiltonian adopted here was first applied 
in IT-NCSM calculations for the ground-state and 
excitation spectra of $^{12}$C in Ref.~\cite{Roth:2011ar}.

\subsection{SRG Evolution}

In the SRG framework the unitary transformation of an operator,
e.g. the Hamiltonian, is formulated in terms of a flow equation 
\beqn
\frac{d}{d\alpha}H_{\alpha}=[\eta_{\alpha},H_{\alpha}] 
 \label{eq:flow}
 \eeqn
with a continuous flow parameter $\alpha$.  The initial condition for
the solution of this flow equation is given by the 'bare' chiral
Hamiltonian. The physics of the SRG evolution is governed by the
anti-hermitian generator $\eta_{\alpha}$.  A specific form widely used
in nuclear physics~\cite{Roth:2010bm,Bogner:2009bt} is given by
\beqn
\eta_{\alpha}=m_N^2 [T_{\text{int}},H_{\alpha}]
\eeqn
where $m_N$ is the nucleon mass and $T_{\text{int}}$ is the intrinsic
kinetic-energy operator.  This generator drives the Hamiltonian
towards a diagonal form in a basis of eigenstates of the intrinsic
kinetic energy, i.e., towards a diagonal in momentum space.

Along with the reduction in the coupling of low-momentum and
high-momentum components by the Hamiltonian, the SRG induces many-body
operators beyond the particle rank of the initial Hamiltonian. In
principle, all the induced terms up to the A-body level are to be
retained in order that the transformation remains unitary and the
spectrum of the Hamiltonian in an exact A-body calculation is
independent of the flow parameter.  In practice we have to truncate
the evolution at a low particle rank (typically, two or three
nucleons), which violates formal unitarity.  In this situation we can
use the flow parameter as a diagnostic tool to quantify the
contribution of omitted many-body terms \cite{Roth:2011ar,RoCa13}.

Throughout this work, we employ the SRG evolution at the three-nucleon
level and neglect four- and multi-nucleon induced interactions.  For
the application in the NCSM it is convenient to solve the flow
equation for the three-body system using a HO Jacobi-coordinate
basis~\cite{Navratil:1999pw,RoCa13}.  The intermediate sums in the
three-body Jacobi basis are truncated at $N_{\max} = 40$ for channels
with $J \leq 5/2$ and ramp down linearly to $N_{\max} = 24$ for $J
\geq 13/2$. Based on this and the corresponding solution of the flow
equation in two-body space (using either a partial-wave momentum- or
harmonic-oscillator representation) we extract the irreducible two-
and three-body terms of the Hamiltonian for the use in A-body
calculations. A detailed discussion of the SRG evolution in the 3N
sector with benchmarks of the truncations involved can be found in
Ref.~\cite{RoCa13}.

\subsection{Computational Aspects of the Many-Body Calculations}

In our many-body calculations, the size of the largest feasible model
space is constrained by the total number of three-body matrix elements
required as well as by the number of many-body matrix elements that
are computed and stored for the iterative Lanczos diagonalization
procedure. Through a $JT$-coupled scheme and an efficient
on-the-fly decoupling during the calculation of the many-body Hamilton
matrix \cite{Roth:2011ar,RoCa13,OrPo13,PoOr13}, the limit arising from
the handling of 3N matrix-elements has been pushed to significantly
larger many-body model spaces. At present, for mid p-shell nuclei the
number of non-zero many-body matrix elements defines the maximum
$N_{\max}$ that can be reached in NCSM calculations.

For the full NCSM calculations we employ the MFDn
code~\cite{DBLP:conf/sc/SternbergNYMVSL08,DBLP:journals/procedia/MarisSVNY10,DBLP:conf/europar/AktulgaYNMV12}
that is highly optimized for parallel computing.  The calculations
were performed on the Cray XE6 Hopper at NERSC, using up to about
100\,TB of memory across 76,320 cores; and on the Cray XK6 Jaguar at
ORNL, using 180\,TB of memory across 112,224 cores, taking about 40
minutes per $\hbar\Omega$-value at $N_{\max}=8$ for 8 converged
eigenvalues. MFDn has been demonstrated to scale well on these
platforms for these types of runs; scaling runs have been performed up
to 261,120 cores on the Cray XK6 Jaguar~\cite{Maris:2013}.

The IT-NCSM calculations are performed with a dedicated code
\cite{RoCa13,Roth:2009cw} that has been developed to accommodate the
specific demands of an importance-truncated calculation in a framework
optimized for parallel performance. Due to the reduction of the
model-space dimension resulting from the importance truncation,
typically by two orders of magnitude, the many-body Hamiltonian matrix
is significantly smaller and the memory needs are drastically
reduced. An IT-NCSM run targeting 8 positive-parity states of $^{12}$C
in an $N_{\max}=8$ space for $\alpha=0.0625\,\text{fm}^4$ and
$\hbar\Omega=20\,\text{MeV}$ takes about 10 hours wall time on 160
nodes on the Cray XE6 Hopper at NERSC and needs a total of 2.5\,TB of
memory for storing the many-body Hamiltonian matrix in the largest
importance-truncated space. This run includes the construction of the
importance-truncated space, the computation of the many-body
Hamiltonian matrix, and the separate solution of the eigenvalue
problems for 15 different values of the importance threshold
$\kappa_{\min}$. Further details on the set-up of the IT-NCSM
calculations are discussed in Sec. \ref{sec:comparisons}.

\section{Results
\label{sec:results}}

\subsection{Excitation Spectra of $^{12}$C
\label{subsec:Excitations}}

\begin{figure}[thb-]
\includegraphics*[width=3in]{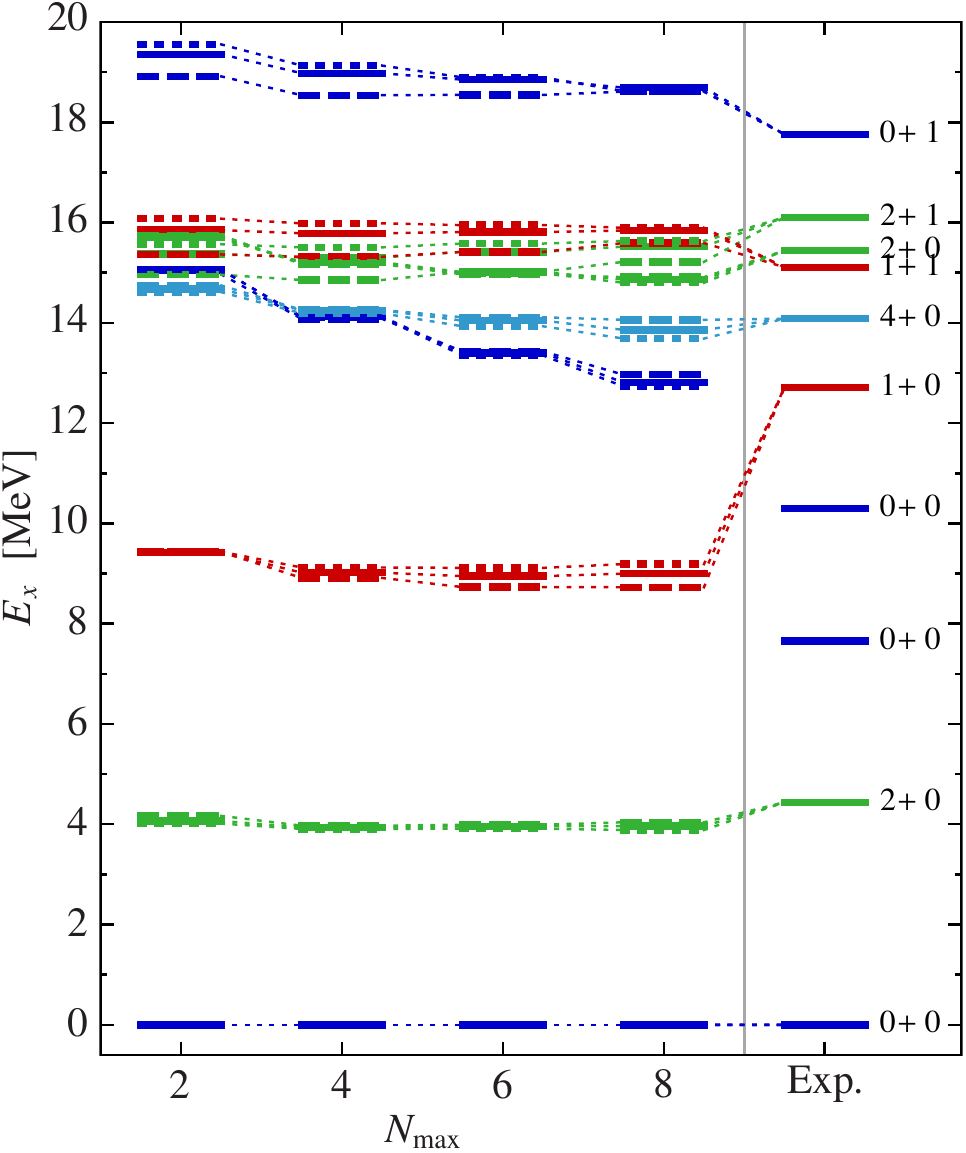}
\caption{(color online) Excitation spectra of $^{12}$C with the chiral NN+3N interactions 
for three different SRG evolution scales 
as a function of $\nmax$ at $\hw=20$ MeV compared with experiment.  
The solid line for each calculated level represents results with $\alpha=0.0625\;\text{fm}^4$.
The long dashed line represents NCSM results with $\alpha=0.08\;\text{fm}^4$ 
and the short dashed line represents NCSM results with $\alpha=0.04\;\text{fm}^4$.  
\label{fig:convergence_SRG}}
\end{figure}

\begin{figure}[thb-]
\includegraphics*[width=3in]{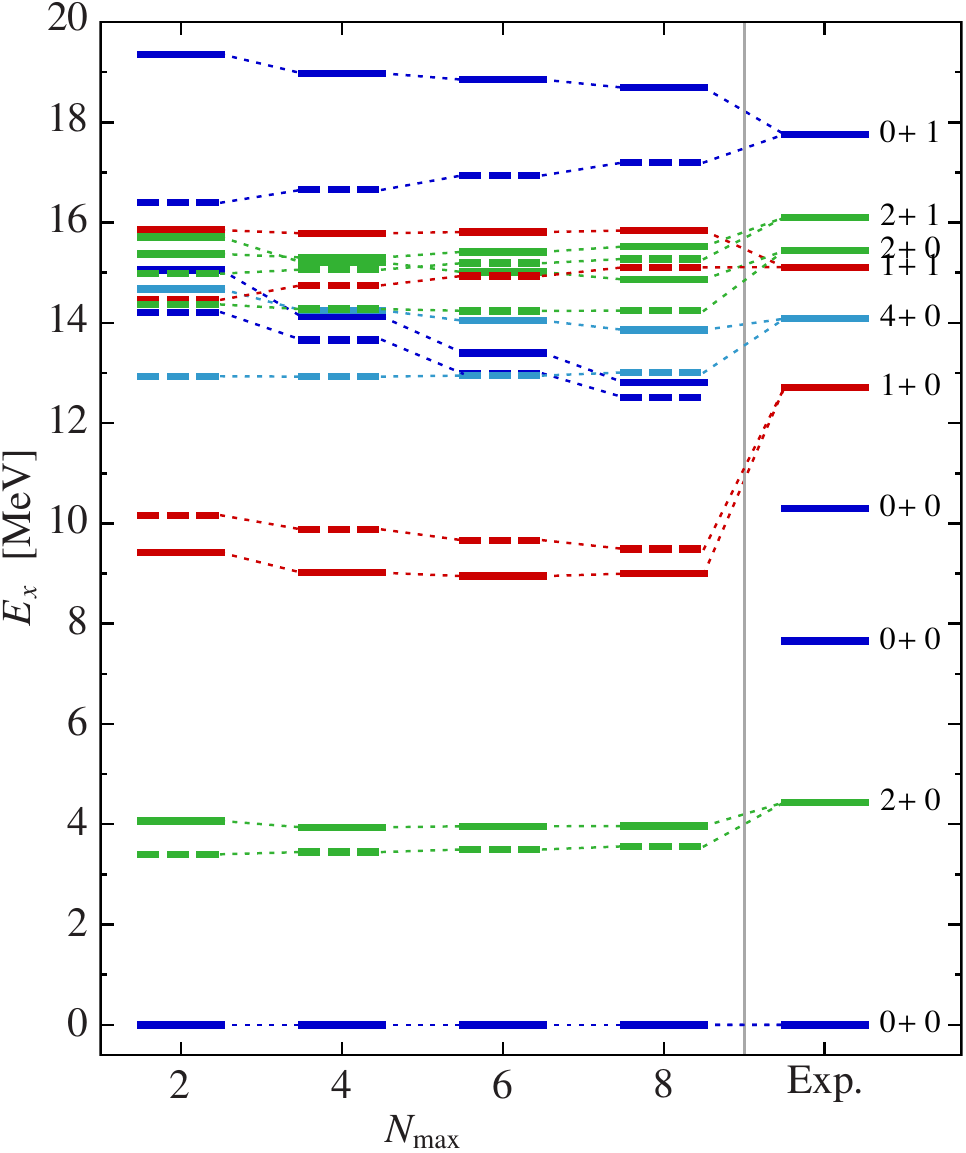}
\caption{(color online) Excitation spectra of $^{12}$C with the chiral NN+3N interactions, 
for two different \hw~values, 
as a function of $\nmax$ at $\alpha=0.0625\,\text{fm}^4$ compared with experiment.  
The solid line for each calculated state represents results with $\hw=20$ MeV.
The long dashed line represents results with $\hw=16$ MeV. 
\label{fig:convergence_hbar_omega}}
\end{figure}

\begin{figure}[thb-]
\includegraphics*[width=3in]{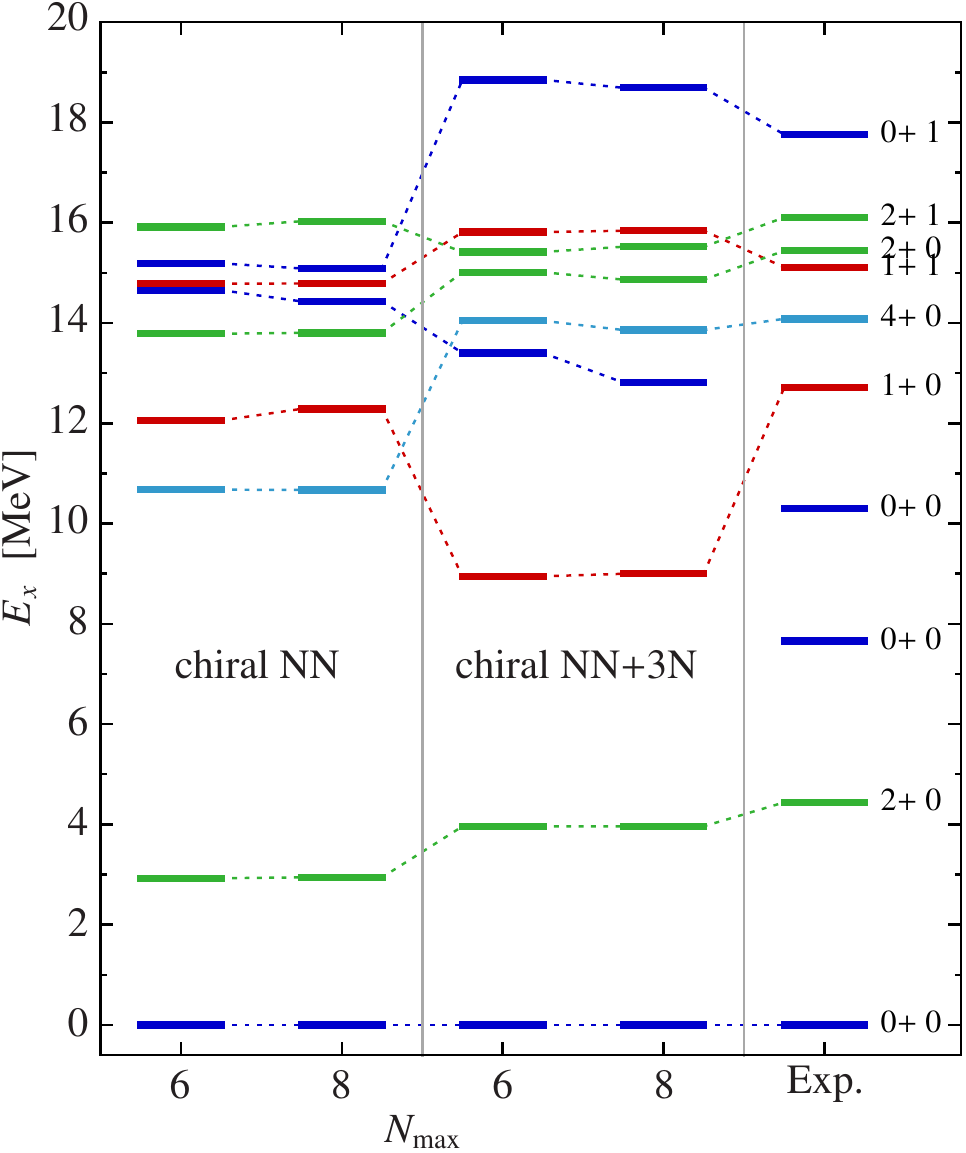}
\caption{(color online) Excitation spectra of $^{12}$C without and with initial chiral 3N interaction for two different \nmax~values, compared with experiment.  
These NCSM results are calculated at $\hw=20$ MeV for flow-parameter $\alpha=0.0625\,\text{fm}^4$.
\label{fig:convergence_bodies}}
\end{figure}

We first investigate the dependence of the excitation spectra of
$^{12}$C on the SRG flow parameter. Starting from the initial chiral
NN+3N interaction, we evolve the Hamiltonian up to a specific
flow-parameter $\alpha$, consistently including two- and three-body
terms, but neglecting SRG-induced four- and multi-nucleon
interactions.  Note that even in cases, where we omit the initial
chiral 3N interaction for comparison purposes, we always include the
SRG-induced 3N terms in our calculations, leading to the so-called
NN+3N-induced Hamiltonian.

Figure \ref{fig:convergence_SRG} shows the behavior of
the excitation energies obtained in the full NCSM with increasing
\nmax~for three values of the SRG flow parameter $\alpha = (0.04,
0.0625, 0.08)\,\text{fm}^4$, corresponding to momentum scales
$\lambda_{\text{SRG}}=\alpha^{-1/4}= (2.24, 2.0, 1.88)\,
\text{fm}^{-1}$.  The spread of converged 
results with the SRG flow parameter will provide an indication of the
relevance of the neglected SRG-induced four- and multi-nucleon
interactions. The absolute ground-state energy starts to
show a non-negligible flow-parameter dependence for $^{12}$C as
discussed in detail in Refs. \cite{Roth:2011ar,RoCa13}.
The excitation energies at fixed $\hbar\Omega$ are 
rather insensitive to the choice of the flow parameter. 
The ground-state and excitation energies, 
as well as additional observables, are provided 
in Table \ref{tab12C_SRG_hw20_lam2.0_NN+3N}.

In Fig.~\ref{fig:convergence_hbar_omega} we display the excitation
spectra of $^{12}$C at two values of \hw~as a function of \nmax. The
spread of the results with \hw~indicates the lack of convergence with
respect to increasing \nmax.  However, the movement of the excitation
energies with increasing \nmax~is consistent with eventual
convergence.

From Figs.~\ref{fig:convergence_SRG} and
\ref{fig:convergence_hbar_omega} one sees that the convergence
patterns are well-enough established to conclude that the
$J^{\pi}T=2^+0$ and $4^+0$ rotational states are reasonably well
reproduced~\cite{Maris:2012ccp}. This is not so surprising in light of
recent successful {\it ab initio} descriptions of collective motion in
light nuclei~\cite{Roth:2010bm,Caprio:2013yp,Dytrych:2013cca}.
However, the $1^+0$ is at least 3 MeV too low as seen in previous
$^{12}$C works using the NCSM with chiral NN+3N
interactions~\cite{Navratil:2007we,Roth:2011ar}.  In addition, we
re-confirm the issue that our basis spaces are insufficient to
reproduce the first excited $0^+0$ state, the Hoyle
state~\cite{Chernykh:2007zz,Neff:2012es}.  Whether the third excited
state at $\nmax=8$, our first excited $0^+0$ state, continues its
downward trend towards the Hoyle state at higher \nmax~values remains
a challenge for the future.

We note that recent lattice simulations with chiral EFT interactions
through N$^2$LO observe the Hoyle state at approximately the correct
excitation energy~\cite{Epelbaum:2011md,Epelbaum:2012qn}.  It will be
interesting to see if the lattice simulated Hoyle state remains in
good agreement with experiment at chiral N$^3$LO and with a range of
lattice spacings.  In addition, it will be interesting to see where
the other low-lying states appear in comparison with experiment.

In order to examine the role of the 3N interaction, we compare in
Fig. \ref{fig:convergence_bodies} the $^{12}$C spectra at SRG
evolution scale $\alpha=0.0625\,\text{fm}^4$ and $\hw=20$ MeV obtained
without and with initial chiral 3N interactions. In both cases the SRG
evolution is performed up to the three-body level: without the initial
3N interaction this leads to the NN+3N-induced Hamiltonian while with
the initial 3N we obtain the NN+3N-full Hamiltonian as used above. We
observe that the impact of the initial chiral 3N interaction is very
different for the various excited states. Whereas the excitation
energies of most states are shifted by about $1$ MeV, some states
exhibit a much stronger sensitivity to the initial 3N interaction.
Among the latter are the first $1^+0$ and the first $0^+1$ states.
The excitation energy of the first $1^+0$ is reduced by more than $3$
MeV by the 3N interaction and the excitation energy of the first
$0^+1$ state is increased by more than $3$ MeV. These large shifts
indicate that these states are strong candidates for sensitive probes
of chiral 3N interactions, particularly for the next-generation
consistent chiral NN+3N Hamiltonians at
N${}^3$LO \footnote{Developments along these lines are ongoing within
  the LENPIC collaboration. See http://www.lenpic/org}. Note, however,
that these excitation energies are not yet converged as seen in the
\nmax\ and \hw-dependence of Fig. \ref{fig:convergence_hbar_omega}.

Another noteworthy effect of including the full 3N interaction seen in
Fig. \ref{fig:convergence_bodies} is to increase the excitation
energies of the lowest rotational excitations, the $2^+0$ and $4^+0$,
by about 30\%.  This increase may be understood as a similar decrease
in the moment of inertia brought about by the increase in binding
energy. Indeed, the ground state rms radius and quadrupole moments are
decreased by the inclusion of the full 3N interaction as discussed
below.

We note that the results at $\nmax=6$ are similar, both in the
locations of excited states and in the changes with the inclusion of
the chiral 3N interaction, with the previous $\nmax=6$ results of
Ref.~\cite{Navratil:2007we}. This similarity is remarkable considering
the different \hw\ values and the different renormalization
schemes---Ref.~\cite{Navratil:2007we} used $\hw=15$ MeV and the OLS
transformation.
 
\subsection{Survey of Observables}
\label{sec:survey}

\begin{table*}[ht]
\begin{ruledtabular}
	\begin{tabular}{c|lll|lll|l}
                                          & \multicolumn{3}{c|}{chiral NN}  &   \multicolumn{3}{c|}{chiral NN+3N} & Experiment \\
	 $N_{\max}$                        & (4,5)     & (6,7)          & (8,9)            & (4,5)    & (6,7)            & (8,9)        &   \\
	\hline
		$E(0_1^+ 0)$ [MeV]                    & -68.123 & -73.483       & -76.617        & -85.756 & -92.182        & -95.761     & -92.161   \\ 
		                                  & -68.123 & -73.544(40) & -76.238(90)  & -85.756 & -92.229(16)  & -95.662(45)  &          \\ 
		$r_p(0_1^+ 0)$ [fm]                            & 2.217   & 2.263          & 2.305           & 2.120   & 2.136          &  2.149         & 2.35(2)   \\ 
		                                            & 2.217  & 2.264(1)       & 2.284(10)    & 2.120   & 2.136(1)     & 2.140(9)       &                \\ 
	$Q (2_1^+ 0)$ [$e$ fm$^2$]        & 4.735  & 5.107           & 5.451           & 3.936   & 4.136          & 4.321           & 6(3)     \\ 
		                                            & 4.735  & 5.129(30)     & 5.191(200)   & 3.936   & 4.155(27)   & 4.232(160)   &               \\ 
	\hline
	$E_{\rm x}(2_1^+ 0)$ [MeV]        & 2.918  & 2.926            & 2.943          & 3.939    & 3.960        & 3.962          & 4.439      \\ 
		                                            & 2.918  & 2.921(6)        & 2.881(12)   & 3.939    & 3.962(4)    & 3.980(19)   &                 \\ 
	$E_{\rm x}(0_2^+ 0)$ [MeV]        & 15.008 & 14.655        & 14.430         & 14.122 & 13.402       & 12.812       &  7.654       \\ 
		                                            & 15.008 & 14.667(15)  & 14.436(26)  & 14.121 & 13.426(16) & 13.066(38) &                \\ 
	$E_{\rm x}(1_1^+ 0)$ [MeV]        & 11.886 & 12.056         & 12.288        & 9.017  & 8.948           & 8.998         & 12.710     \\ 
		                                            & 11.886 & 12.050(13)  & 12.116(23)  & 9.018  & 8.951(9)       & 8.891(20)   &                  \\ 
	$E_{\rm x}(4_1^+ 0)$ [MeV]        & 10.704 & 10.676        & 10.670        & 14.250 & 14.044         & 13.860       & 14.083     \\ 
		                                            & 10.704 & 10.682(7)    & 10.703(10) & 14.250 & 14.052(8)     & 14.015(33) &                \\ 
	$E_{\rm x}(1_1^+ 1)$ [MeV]        & 14.819 & 14.786        & 14.788        & 15.787 & 15.812        & 15.841       & 15.110     \\ 
		                                            & 14.819 & 14.774(15)  & 14.712(32) & 15.787 & 15.820(8)   & 15.833(23) &                \\ 
	$E_{\rm x}(2_2^+ 0)$ [MeV]        & 13.834 & 13.787         & 13.803       & 15.206 & 15.012        & 14.865       & (15.44)     \\ 
		                                            & 13.834 & 13.784(9)    & 13.719(10) & 15.206 & 15.017(4)   & 14.950(29) &                \\ 
	$E_{\rm x}(2_1^+ 1)$ [MeV]        & 15.781 & 15.916        & 16.030        & 15.304 & 15.416        & 15.521       & 16.106      \\ 
		                                            & ------     & ------            & ------            & 15.305 & 15.419(9)    & 15.430(40) &                \\ 
	$E_{\rm x}(0_1^+ 1)$ [MeV]        & 15.359 & 15.189        & 15.088         & 18.978 & 18.850        & 18.691        & 17.760      \\ 
		                                            & 15.359 & 15.182(9)    & 15.020(40)   & ------     & ------            & ------           &                \\ 
	\hline
	$E(3_1^- 0)$ [MeV]                & -55.010  &  -61.249     & ------            & -70.460 & -77.336         &   ------       &   -82.520     \\ 
		                              & -55.010 & -61.182(150) & -62.883(400)  & -70.460  & -77.464(120) &  -79.961(400) &          \\ 
	$E_{\rm x}(3_1^- 0)$ [MeV]        & 13.113  & 12.234          & ------                & 15.296  & 14.846          & ------             & 9.641       \\ 
		                              & 13.113  & 12.362(170)  & 13.355(450) & 15.297  & 14.765(150)  & 15.701(450) &                \\ 
	$E_{\rm x}(1_1^- 0)$ [MeV]        & 16.079  & 15.079           & ------              & 17.703  & 17.089          & ------             & 10.844      \\ 
		                              & 16.079  & 15.217(170)  & 15.937(450) & 17.703  & 16.999(150)  & 17.688(450) &                 \\ 
	$E_{\rm x}(2_1^- 0)$ [MeV]        & 17.081  & 16.182           & ------                & 17.937  & 17.429          & ------              & 11.828     \\ 
		                              & 17.080  & 16.304(170)   & 17.059(450) & 17.937  & 17.305(150)  & 17.905(450) &       \\ 
    $E_{\rm x}(4_1^- 0)$ [MeV]        & 16.944  &  16.122           & ------                      & 19.030  & 18.579          & ------              & (13.352)   \\ 
		                              & 16.943  & 16.282(170)   & 17.348(450) & 19.030  & 18.508(150)  & 19.482(450) &       \\ 
	\hline	
	B(E2;$2_1^+0 \rightarrow 0_1^+0$) [$e^2$ fm$^4$] & 5.001  & 5.834         & 6.689          & 3.558   & 3.885         & 4.210          & 7.59(42)     \\ 
		                                                  & 5.001  & 5.844(18)   & 6.504(90)    & 3.558   & 3.894(8)    & 4.080(75)    &                 \\ 
	B(M1;$1_1^+0 \rightarrow 0_1^+0$) [$\mu_N^2$] & 0.0032 & 0.0030        & 0.0030       & 0.0080 & 0.0079      & 0.0078        & 0.0145(21) \\ 
		                                                  & 0.0032 & 0.0030(1)   & 0.0032(2)   & 0.0079 & 0.0078(1)   & 0.0082(3)   &                 \\ 
	B(M1;$1_1^+1 \rightarrow 0_1^+0$) [$\mu_N^2$] & 0.388  & 0.343       & 0.304             & 1.157   & 1.139         & 1.109         & 0.951(20)   \\ 
		                                                  & 0.388  & 0.343(1)  & 0.329(6)         & 1.157   & 1.135(5)     & 1.143(36)   &                \\ 
	B(E2;$2_1^+1 \rightarrow 0_1^+0$) [$e^2$ fm$^4$] & 0.308  & 0.293       & 0.241             & 0.437   & 0.442         & 0.436          & 0.65(13)     \\ 
		                                                 & ------    & ------         & ------               & 0.437   & 0.440(7)    & 0.444(18)    &                \\ 
	\end{tabular}

\end{ruledtabular}
\caption{Calculated and experimental total energies $E$, excitation
  energies $E_{\rm x}$, point-proton rms radii $r_p$, quadrupole
  moments $Q$, as well as E2 transitions B(E2), and M1 transitions
  B(M1) of $^{12}$C. The first 3 columns correspond to results for the
  initial chiral NN interaction (still including SRG-induced 3N-terms)
  while the next 3 columns correspond to chiral NN+3N interaction
  using an SRG evolution scale $\alpha=0.0625\,\text{fm}^4$
  ($\lambda_{\text{SRG}}=2.0$ fm$^{-1}$) and $\hbar\Omega=20$ MeV.
  Columns of theoretical results are labelled by pairs of natural and
  unnatural parity basis spaces characterized by their $N_{\max}$
  values.  The first row of each observable is obtained with the NCSM
  while the second row is obtained from the IT-NCSM. The uncertainty
  extracted from the threshold extrapolation of the IT-NCSM results as
  discussed in the text are quoted in parenthesis; for
  $N_{\max}=(4,5)$ the full space was used.  The experimental values
  are taken from Ref.~\cite{AjSe90,DeDe87,VeEs83}.}
\label{tab12C_SRG_hw20_lam2.0_NN+3N}
\end{table*}

In addition to the spectra shown in the figures above, we present in Table
\ref{tab12C_SRG_hw20_lam2.0_NN+3N} the ground-state energy, selected
excitation energies and a survey of electromagnetic observables in
$^{12}$C for one choice of SRG flow-parameter,
$\alpha=0.0625\,\text{fm}^4$, and one choice of HO basis frequency,
$\hw=20\mev$.  While many cases were generated to perform our
systematic survey and prepare the figures, we have chosen this one
representative case, with a moderate value of the SRG evolution scale,
to present in more detail. We tabulate these results in order to
stimulate detailed comparisons with other methods and other
Hamiltonians. In addition, we specify the IT-NCSM results for the
benchmark comparison discussed in detail in the
Sec. \ref{sec:comparisons}.

\begin{figure}[tbh-]
\includegraphics*[width=1.08\columnwidth]{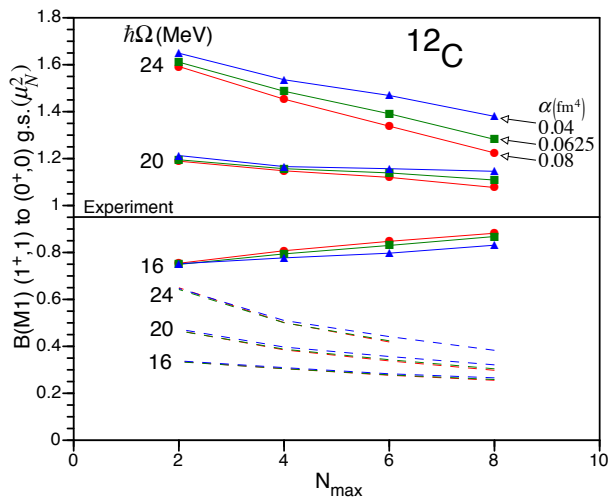}
\caption{(color online) Reduced magnetic dipole transition matrix
  element from the $1^+1$ to the ground state of $^{12}$C (in
  units of $\mu_N^2$) as a function of \nmax\ at three different
  SRG evolution scales and three different HO basis frequencies. The
  solid and dashed lines present the results obtained with and without
  the initial chiral 3N interaction, respectively.
\label{fig:BM1_C12}}
\end{figure}

\begin{figure}[tbh-]
\includegraphics*[width=1.08\columnwidth]{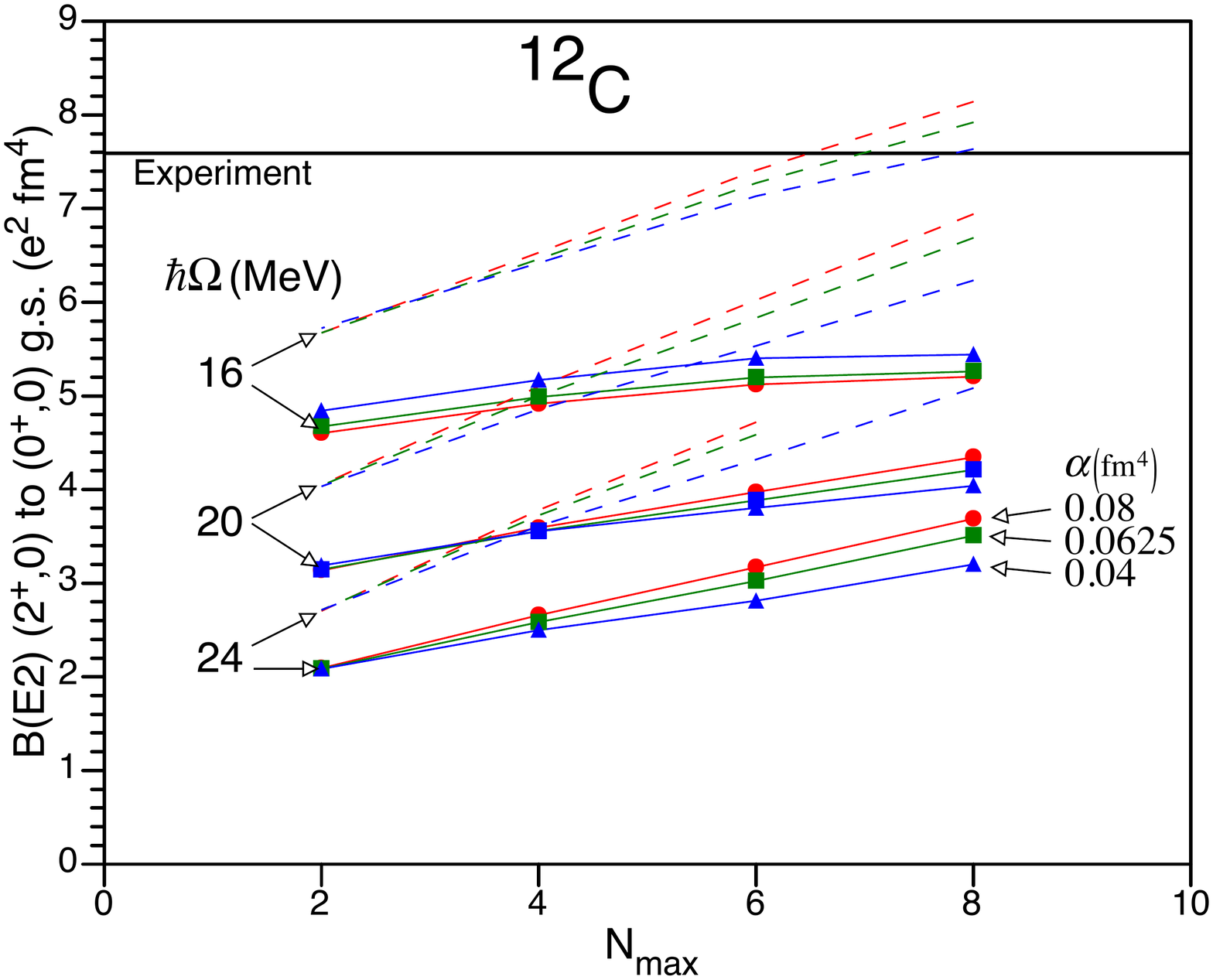}
\caption{(color online) Reduced electric quadrupole transition matrix
  element from the $2^+0$ to the ground state of $^{12}$C (in units of
  $e^2$ fm$^4$) as a function of \nmax\ at three different SRG
  evolution scales and three different HO basis frequencies.  The
  solid and dashed lines present the results obtained with and without
  the initial chiral 3N interaction, respectively.
\label{fig:BE2_C12}}
\end{figure}

In order to more completely understand the basis space (\nmax, \hw)
dependence as well as the flow-parameter dependence of two selected
electromagnetic observables, we present these results as a
function of \nmax\ in Figs.~\ref{fig:BM1_C12} and \ref{fig:BE2_C12}.

The example of the B(M1) from the $1^+1$ to the ground state has
previously been identified as receiving about a factor of three
enhancement when 3N interactions are
included~\cite{Hayes:2003ni}. This earlier work used the
Tucson-Melbourne TM'(99) interaction~\cite{Coon:2001pv} in NCSM
calculations up through $\nmax=6$ to establish this enhancement.  This
enhancement has been confirmed with chiral NN+3N interactions in NCSM
calculations also through $\nmax=6$ using the OLS renormalization
approach~\cite{Navratil:2007we}.  In Fig.~\ref{fig:BM1_C12} we
reconfirm this result with chiral NN+3N interactions up through
$\nmax=8$ and show the sensitivity to the SRG flow-parameter and to
the basis-space parameters (\nmax,\hw).  Clearly, these dependences
are weak enough that the general conclusion remains---this B(M1) is
strongly enhanced by 3N interactions and the amount of enhancement is
roughly independent of the adopted Hamiltonian.

Contrasting the favorable convergence picture for the B(M1), other
observables that are sensitive to the extent of the wavefunction, such
as rms radii, quadrupole moments and B(E2)'s, are not well converged
(see Table \ref{tab12C_SRG_hw20_lam2.0_NN+3N}).  Of course, the radial
extent is sensitive to the binding energy relative to the first
threshold which is the 3$\alpha$ threshold at about 7 MeV
experimentally.  This allows an intuitive interpretation of our
results for the B(E2) from the lowest $2^+0$ to the ground state, see
Fig. \ref{fig:BE2_C12}.  One of the effects of the inclusion of the
chiral 3N interaction is that the 3$\alpha$ threshold is pushed to
higher excitation energies.  Without the chiral 3N interaction, both,
$^4$He and $^{12}$C are underbound and we find the 3$\alpha$ threshold
at too low excitation energies. The inclusion of the 3N interaction
increases the binding energy of $^4$He and of $^{12}$C such that the
3$\alpha$ threshold is pushed to higher excitation energies. Not
surprisingly, the changes in the B(E2)s are well correlated with the
changes in the binding and threshold energies. Also, the changes in
the B(E2)s are well correlated with the changes in the ground state
rms radius and the quadrupole moment of the first $2^+0$.

One has to keep in mind, however, that two components are still
missing in the present, state-of-the-art calculations of
electromagnetic observables: First, we should transform the
electromagnetic operators consistently with the Hamiltonian in the SRG
evolution. However, so far we only employ the bare operators. Second,
we should include the two-body currents derived in chiral
EFT. However, so far we only employ the one-body part. These two
corrections are not likely to affect the qualitative discussion
present here, but they will play a role in future precision
calculations of electromagnetic observables.

\subsection{Negative Parity States}
\label{sec:Negparity}

\begin{figure}[tbh-]
\includegraphics*[width=1.08\columnwidth]{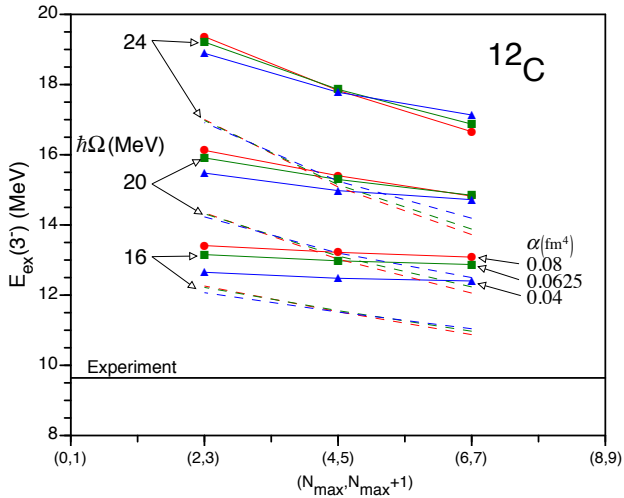}
\caption{(color online) Excitation energy of the $3^-0$ in $^{12}$C as
  a function of (\nmax,\nmax\ +1) without (dashed lines) and with (solid lines) 
  initial chiral 3N interaction at
  three different SRG evolution scales and three different HO basis
  frequencies. The unnatural parity states are computed 
  at \nmax\ +1 while the corresponding excitation energy is calculated with
  respect to the ground state at \nmax\ so that a pair of basis
  spaces defines each point in this plot.
\label{fig:Eex_3m_C12}}
\end{figure}

\begin{figure}[thb-]
\includegraphics*[width=0.9\columnwidth]{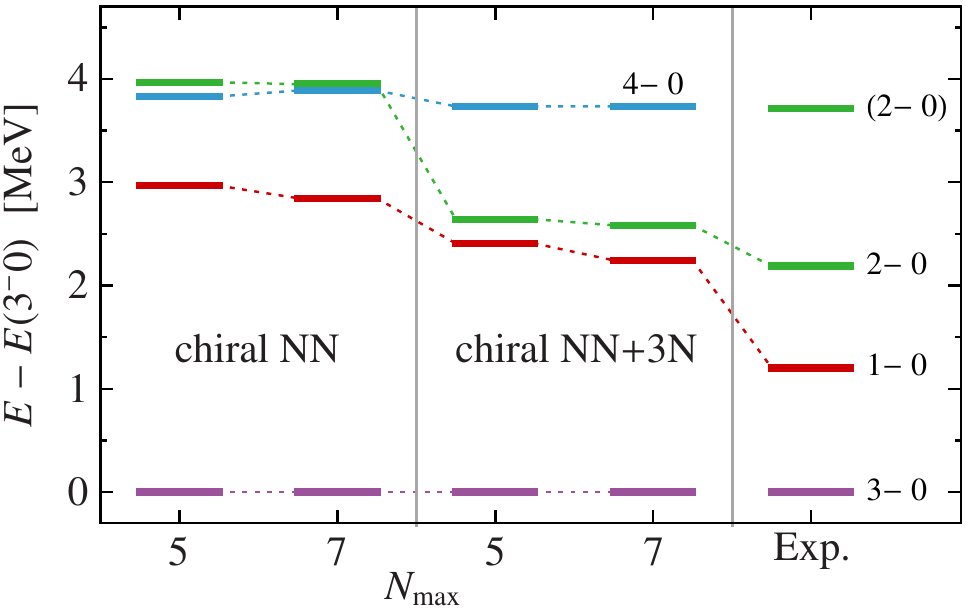}
\caption{(color online) Negative parity excitation spectra of $^{12}$C
  obtained without and with initial chiral 3N interaction for two
  different \nmax~values, compared with experiment.  These NCSM
  results are calculated at $\hw=20$ MeV for flow parameter
  $\alpha=0.0625\,\text{fm}^4$.
\label{res_spectrum_unn}}
\end{figure}

Finally, we present in Figs.~\ref{fig:Eex_3m_C12} and
\ref{res_spectrum_unn} results for the lowest excited states with
negative parity in $^{12}$C.  Figure \ref{fig:Eex_3m_C12} displays an
array of results for the $3^-0$ state from our current investigation,
covering the three frequencies $\hw=16, 20,$ and $24\,\text{MeV}$ and
the three SRG flow parameters $\alpha=0.04, 0.0625,$ and
$0.08\,\text{fm}^4$. Here we show the energy difference of the $3^-0$
state and the $0^+0$ ground state obtained in $N_{\max}+1$ and
$N_{\max}$ spaces, respectively.

There is a sizable spread of the excitation energy of the $3^-0$ state
in Fig.~\ref{fig:Eex_3m_C12} with both frequency and $N_{\max}$,
indicating a slow convergence compared to the typical positive-parity
states, both with and without the initial 3N interaction.
Nevertheless, our results indicate that the initial chiral 3N
interaction increases the excitation energy of the $3^-0$ state by a
few MeV.

The excitation spectra of the lowest negative parity states relative
to the $3^-0$ state are better converged, see
Fig.~\ref{res_spectrum_unn}.  Without the chiral 3N interaction the
lowest $1^-0$ state and even more significantly the $2^-0$ states are
too high in the negative-parity spectrum. The chiral 3N interaction
reduces the excitation energy of both the $1^-0$ and $2^-0$ states,
and brings them in better agreement with the experimental data.  Our
calculations also indicate that the fourth negative parity state is a
$4^-0$ state, in agreement with the $J^\pi$ assignment suggested by
Millener~\cite{Millener:comment}.

\section{Benchmark of IT-NCSM with NCSM
\label{sec:comparisons}}

Apart from the discussion of the spectroscopy and other observables of
$^{12}$C obtained with chiral NN+3N interactions, a second main goal
of the present work is to present a benchmark comparison between NCSM
and IT-NCSM for ground- and excited-state energies and electromagnetic
observables. To this end, Table \ref{tab12C_SRG_hw20_lam2.0_NN+3N}
contains the numerical results from NCSM and IT-NCSM calculations for
our selected $^{12}$C observables in a pair-wise comparison.

The general setup of the IT-NCSM calculations presented here is as
follows. For basis spaces up to $\nmax=(4,5)$ we use the IT-NCSM code
\cite{RoCa13,Roth:2009cw} for full NCSM calculations. The energies for
these full-space runs agree to within 1 keV and the electromagnetic
observables to within 0.1\% with the NCSM results obtained with the
MFDn code. This establishes a baseline for the numerical precision of
the two independent codes, which use the same $JT$-coupled NN and 3N
matrix elements as input.

Beginning at $\nmax=6$, the IT-NCSM calculations involve the
importance truncation and threshold extrapolation. A detailed
discussion of the IT-NCSM can be found in
Refs.~\cite{RoCa13,Roth:2009cw}. For the positive-parity spectrum,
e.g., we target the 8 lowest eigenstates. For each of them we define a
reference state by using the corresponding eigenstate obtained in the
next-smaller model space and imposing a reference threshold
$C_{\min}=2\times10^{-4}$ that eliminated all components with
amplitudes below this threshold. This value is expected to be
sufficiently small not to affect the final results, for a detailed
analysis see Ref.~\cite{RoCa13}. These reference states enter into the
importance measure used to identify the relevant basis states for the
description of any one of the 8 target states, i.e., if the importance
measure with respect to at least one reference state is above the
importance threshold $\kappa_{\min}$ the basis state is kept. We
employ a sequence of importance thresholds $\kappa_{\min}=\{3, 3.5,
4,...,10 \}\times 10^{-5}$ and solve for the eigenvalues within each
of the importance-truncated model spaces separately. Based on the
energies and observables obtained for the different
importance-truncated spaces we perform an extrapolation
$\kappa_{\min}\to0$. We use a third-order polynomial fit to the
results for the full range of importance thresholds with equal
weights. The uncertainty of the threshold extrapolation is quantified
by changing the order of the polynomial by $\pm 1$ and by excluding
the results of the lowest and the lowest two threshold values. The
uncertainties quoted in Table \ref{tab12C_SRG_hw20_lam2.0_NN+3N} are
the standard deviations obtained for this set of extrapolations.

The results of the full NCSM in these larger basis spaces serve as
important benchmarks for the IT-NCSM.  In general we observe a very
good agreement of the IT-NCSM results with the full NCSM with
deviations below 1\% for almost all cases.  Beyond assessing this
general agreement, the full NCSM results provide a unique opportunity
to test the reliability of the uncertainty estimates obtained from the
threshold extrapolation protocol discussed above.  
It should be noted that we do not account for the numerical
uncertainty in the NCSM result used as the benchmark.  For energies
this NCSM uncertainty is expected to be about 1 keV; for
electromagnetic observables the NCSM uncertainty has not been
quantified previously, but based on the excellent agreement of the
full-space results for $N_{\max}=(4,5)$ with an independent code we
expect uncertainties of the order of the last quoted digit.

Considering all NCSM and IT-NCSM pairs of results in Table
\ref{tab12C_SRG_hw20_lam2.0_NN+3N}, we observe that the IT-NCSM agrees
within the quoted uncertainty with the NCSM result in 60\% of the
cases.  From this observation one might conclude that the procedure
used to quantify the IT-NCSM uncertainties is reasonable and may be
interpreted in a similar way as a statistical standard deviation.
However, there are specific patterns in the size of the estimated
uncertainties and the agreement with the full NCSM results.

\begin{figure}[thb-]
\includegraphics*[width=3in]{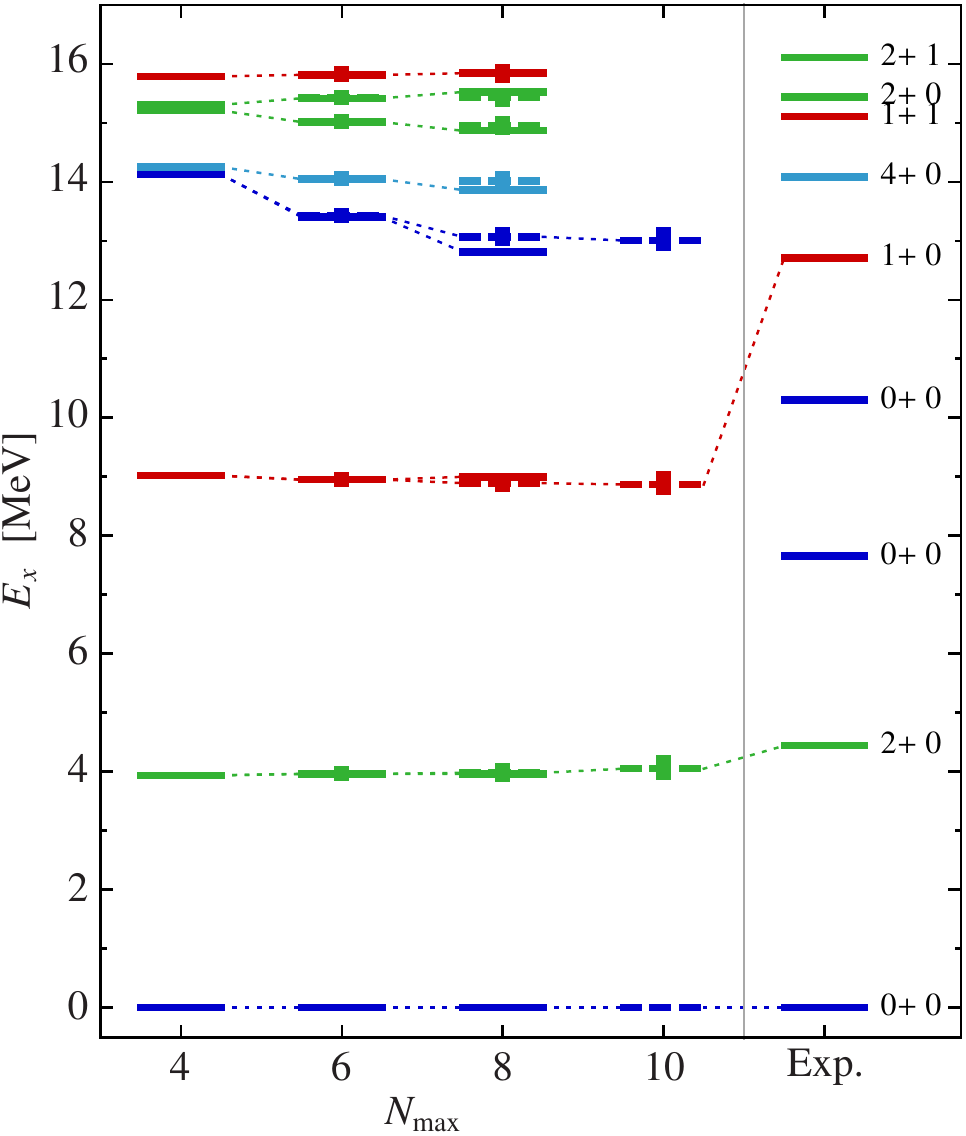}
\caption{(color online) Excitation spectra of $^{12}$C with the chiral
  NN+3N interactions, obtained with the NCSM and the IT-NCSM, as a
  function of \nmax, and compared with experiment.  The solid lines
  represent the NCSM result and dashed lines represent the IT-NCSM
  results, with boxes indicating the typical threshold-extrapolation
  uncertainties. These results are calculated at $\hw=20$ MeV with the
  SRG evolution scale $\alpha=0.0625\,\text{fm}^4$.  For $\nmax=10$,
  only IT-NCSM calculations targeting the lowest four eigenstates are
  currently available.
\label{fig:convergence_benchmark}}
\end{figure}

\begin{figure}[thb-]
\includegraphics*[width=0.9\columnwidth]{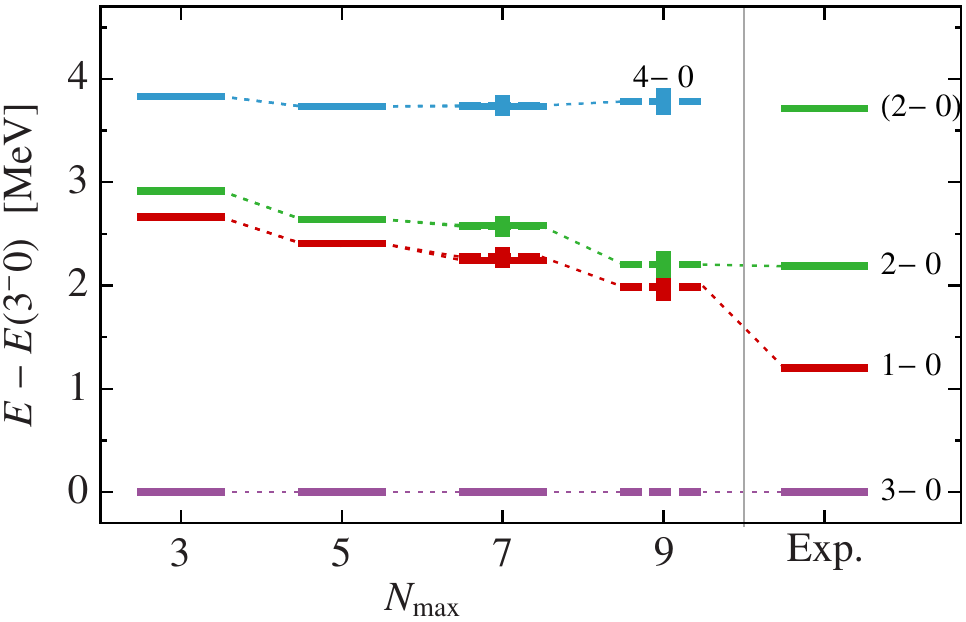}
\caption{(color online) Excitation of the negative parity spectra 
of $^{12}$C with respect to the lowest $3^-0$ state using the chiral
  NN+3N interaction, obtained with the NCSM and the IT-NCSM, as a
  function of \nmax, and compared with experiment.  The solid lines
  represent the NCSM result and dashed lines represent the IT-NCSM
  results.  These results are calculated at $\hw=20$ MeV with the SRG
  evolution scale $\alpha=0.0625\,\text{fm}^4$.  For $\nmax=9$, only
  IT-NCSM calculated eigenstates are currently available.
\label{fig:convergence_negative}}
\end{figure}

For the excitation energies of the positive-parity states the
estimated IT-NCSM uncertainties resulting from the threshold
extrapolation are below $50\,\text{keV}$ and the majority of the
IT-NCSM results agree with the full NCSM within the estimated
uncertainty, though the fraction of cases showing an agreement 
within the uncertainties decreases 
significantly for $N_{\max}=8$ compared to $N_{\max}=6$.  

This is illustrated in Fig.~\ref{fig:convergence_benchmark}, where we
display the excitation energies of the positive-parity states obtained
in the full NCSM and with the IT-NCSM; the estimated uncertainties of
the IT-NCSM are indicated by the boxes.  The dashed bars representing
the IT-NCSM results almost always agree within uncertainties with the
solid bars representing the full NCSM.  The only case where the
difference is more pronounced is the excited $0^+0$. The atypical
\nmax\ dependence of this state already hints at a complicated
structure of the wave function which is dominated by small
components---evidently this represents a more difficult situation for
the importance truncation and threshold extrapolation.  For
completeness we also show excitation energies at $\nmax=10$, which
were obtained in an IT-NCSM calculation targeting the four lowest
eigenstates.
  
For the excitation energies of the negative-parity states relative to
the $3^-0$ state, as shown in Fig. \ref{fig:convergence_negative}, the
agreement of the IT-NCSM and the full NCSM is equally good. Based on
the direct threshold extrapolation of the excitation energies within
the negative-parity space, the uncertainties of the IT-NCSM energies
are comparable to the uncertainties of the positive-parity excitation
energies. Note, however, that the uncertainties of excitation energies
of the negative-parity states relative to the positive-parity ground
state, as reported in Table \ref{tab12C_SRG_hw20_lam2.0_NN+3N}, are
significantly larger. This results from the larger uncertainties of
the threshold extrapolations for the absolute energies of the $3^-0$
and the $0^+0$ states needed to determine the offset of the
negative-parity with respect to the positive-parity spectrum. The
uncertainties in this offset induce sizeable systematic uncertainties
in the excitation energies of the negative-parity states, as seen in
Table \ref{tab12C_SRG_hw20_lam2.0_NN+3N}.

For radii and electromagnetic observables the threshold extrapolations
typically produce larger error bars, particularly for long-range
observables like the radii or quadrupole moments and
transitions. Nevertheless, even for these observables, the results in
Table \ref{tab12C_SRG_hw20_lam2.0_NN+3N} show that the NCSM and
IT-NCSM results fall within the quoted IT-NCSM uncertainty in the
majority of cases.

There is a systematic trend in uncertainties of the IT-NCSM results
when going from $N_{\max}=(6,7)$ to $N_{\max}=(8,9)$. First of all,
the uncertainty estimates increase with increasing $N_{\max}$. This is
due to the fact that the IT-NCSM space covers a smaller fraction of
the complete $N_{\max}$ space so that the threshold extrapolation has
to account for the contribution of a larger fraction of discarded
basis states. Second, the fraction of cases in which the IT-NCSM
agrees with the NCSM within the uncertainties is reduced for
$N_{\max}=(8,9)$. This might be explained by uncertainties that are
not accounted for by the threshold extrapolation and uncertainty
quantification protocol. An example are inaccuracies resulting from
building the importance-truncated space for $N_{\max}=(8,9)$ on
reference states that already result from an importance-truncated
$N_{\max}=(6,7)$ calculation---the uncertainties inherited from the
$N_{\max}=(6,7)$ states and the additional reference threshold $C_{\min}$ 
are not yet accounted for by the $N_{\max}=(8,9)$
uncertainty estimate. Since a numerical propagation of these
uncertainties is computationally expensive, one might consider other
threshold extrapolation schemes that are robust in this respect. A
promising candidate is a threshold extrapolation based on the energy
variance \cite{ZhNo04,Abe:2012wp} and studies along these lines are in
progress.

These benchmark comparisons show that the intrinsic uncertainty
estimates extracted from the threshold extrapolation provide a
suitable guideline for the accuracy of the IT-NCSM results. 
However, one has to keep in mind that the estimates 
do not capture the accumulation of uncertainties 
throughout a sequence of importance truncated calculations 
with increasing $N_{\max}$. The
relative size of the uncertainties depends on the observable and the
structure of the states.  If the resulting uncertainty appears too
large for a specific application, one may elect to decrease the
importance thresholds, which is guaranteed to improve the results and
reduce the extrapolation uncertainties.

\section{Summary and Conclusions}
\label{sec:conclusions}

We have presented \abinit NCSM and IT-NCSM calculations of $^{12}$C
using SRG-evolved chiral NN+3N Hamiltonians.  Both, spectra and
electromagnetic properties are examined as a function of the SRG
flow-parameter as well as a function of the model-space parameters
(\nmax,\hw).  We have extended previous investigations with the same
Hamiltonian to larger model spaces and to a larger set of observables.
Furthermore, we have benchmarked the IT-NCSM with the NCSM in model
spaces where the latter is feasible.

For most low-lying positive-parity states the excitation energies are
reasonably well converged, though noticeable exceptions are the first
excited $0^+0$ state (the Hoyle state) and the first $0^+1$ state.
Indeed, it is known that in order to converge the Hoyle state one
needs significantly larger model spaces.  Electromagnetic observables
such as magnetic moments and M1 transition strengths are also
reasonably well converged, but quadrupole moments and E2 transitions
are not yet converged, even in the largest model spaces that we have
considered here.  This should however not be surprising, since the E2
operator is a long-range operator, and hence convergence is
notoriously slow in a HO basis.

The comparison of our theoretical spectra with experiment reveals some
remarkable points.  For most of the positive-parity states the
excitation energies obtained without the initial chiral 3N interaction
are in good qualitative agreement with experiment.  Typically the
agreement is improved by including the chiral 3N interaction, in
particular for the rotational excitations, the lowest $2^+0$ and
$4^+0$ states.  Also the $0^+1$ state is very sensitive to the 3N
interaction, and in better agreement with experiment than without the
3N interaction.  A surprising exception is the $1^+0$ state.  It is in
good agreement with experiment without the chiral 3N interaction.
However, with the chiral 3N interaction the excitation energy is
pushed about 4\,MeV below the experimental value.

The excitation energies of the lowest excited negative-parity states
with respect to the $3^-0$ state (the lowest negative-parity state)
are also reasonably well converged.  However, the negative-parity
states converge slower than the positive-parity states in terms of
absolute energies, and hence the excitation energies of the
negative-parity states are not converged with respect to the ground
state.  Nevertheless, it appears that the dominant effect of the
chiral 3N interaction on the negative-parity states is an overall
upward shift of the states with respect to the positive-parity ground
state.
  
The excitation energies of the $1^+0$ state and the $0^+1$ state
represent valuable test cases for next-generation (chiral)
Hamiltonians.  The failure of the present chiral NN at N$^3$LO plus 3N
interaction at N$^2$LO to quantitatively capture the physics of these
states represents a challenge for improved chiral interactions, in
particular the 3N interaction at N$^3$LO~\cite{BeEp08,BeEp11}.
Further detailed investigations into the structure of these states and
their sensitivity to different existing chiral NN+3N interactions are
in progress.

In terms of enlarging the model space, which may be necessary in order
to address these issues, there are alternatives to the straightforward
but challenging task of enlarging the HO basis itself.  In particular,
it may be more fruitful to adopt another basis with improved infrared
properties such as the Coulomb-Sturmian
basis~\cite{Keister:1996bd,Caprio:2012rv,Caprio:2012as}.
Alternatively, it may be more efficient to remain within the HO basis
(thereby preserving factorization of the center-of-mass motion) but
selecting symmetry-adapted basis spaces such as those recently
advocated for light nuclei~\cite{Dytrych:2013cca} and especially for
$^{12}$C~\cite{Dreyfuss:2012us,Dytrych:2014}.  Another avenue is the
explicit treatment of clusters and their relative motion in the NCSM
with continuum (NCSMC) that was recently formulated and successfully
applied to the description of the unbound nucleus $^7$He
\cite{BaNa13b}.

\begin{acknowledgments}

This work was supported in part by the US National Science Foundation
under Grant No. PHY--0904782, the US Department of Energy (DOE) under
Grant No.~DE-FG02-87ER40371 and DESC0008485 (SciDAC-3/NUCLEI), by the
Deutsche Forschungsgemeinschaft through contract SFB 634, by the
Helmholtz International Center for FAIR (HIC for FAIR) within the
LOEWE program of the State of Hesse, and the BMBF through contract
06DA7047I.  This work was supported partially through GAUSTEQ (Germany
and U.S. Nuclear Theory Exchange Program for QCD Studies of Hadrons
and Nuclei) under contract number DE-SC0006758.  A portion of the
computational resources were provided by the National Energy Research
Scientific Computing Center (NERSC), which is supported by the US DOE
Office of Science, and by an INCITE award, "Nuclear Structure and
Nuclear Reactions", from the US DOE Office of Advanced Scientific
Computing.  This research also used resources of the Oak Ridge
Leadership Computing Facility at ORNL, which is supported by the US
DOE Office of Science under Contract DE-AC05-00OR22725. Further
resources were provided by the computing center of the TU Darmstadt
(lichtenberg), the J\"ulich Supercomputing Centre (juropa), and the
LOEWE-CSC Frankfurt.

\end{acknowledgments}


%

\end{document}